\documentclass{aa}
\usepackage[utf8]{inputenc}
\usepackage{natbib}
\usepackage{amsmath}
\usepackage{graphicx}
\usepackage{color}

\def\gr{$\gamma$-ray}
\def\aap{A\&A}

\def\apjl{AP.J.Lett.}
\usepackage[normalem]{ulem}

\begin{document}

\title{Mapping large-scale diffuse gamma-ray emission in the 10-100 TeV band with Cherenkov telescopes }
\author{
A. Neronov \inst{1,2} \and
D. Semikoz \inst{1} }
 \date{}
\institute{APC, University of Paris, CNRS/IN2P3, CEA/IRFU,  10 rue Alice Domon et Leonie Duquet, Paris, France\and
Astronomy Department, University of Geneva, Ch. d'Ecogia 16, 1290, Versoix, Switzerland}

\label{firstpage}

\abstract
{
Measurement of diffuse \gr\ emission from the Milky Way with  Imaging Atmospheric Cherenkov  Telescopes (IACT) is difficult because of  the high level of charged cosmic ray background and the small field of view. 
 }
{
We show that such a measurement is nevertheless possible in the energy band 10-100 TeV.} 
{
The minimal  charged particle background for IACTs is achieved by selecting the events  to be used for the analyses of the cosmic ray electrons. Tight cuts on the event quality  in these event selections allow us to obtain a sufficiently low background level to allow measurement of the diffuse Galactic \gr\ flux  above 10 TeV. We calculated the sensitivities of different types of IACT arrays for the  Galactic diffuse emission measurements and compared them with the diffuse \gr\ flux from different parts of the sky measured by the Fermi Large Area Telescope below 3 TeV and with the astrophysical neutrino signal measured by IceCube telescope.
}{ 
 We show that deep exposure of existing IACT systems is sufficient for detection of the diffuse flux from all the Galactic Plane up to Galactic latitude $|b|\sim 5^\circ$. The Medium Size Telescope array of the CTA will be able to detect the diffuse flux up $30^\circ$ Galactic latitude. Its sensitivity will be sufficient for detection of the $\gamma$-ray counterpart of the Galactic component of the IceCube astrophysical neutrino signal above 10 TeV.   We also propose that a dedicated IACT system composed of small but  wide-field-of-view telescopes could be used to map the 10-100 TeV diffuse \gr\ emission from across the whole sky. 
}
{
Detection and detailed study of diffuse Galactic \gr\ emission in the previously unexplored 10-100 TeV energy range is possible with the  IACT technique. This is important for identification  of the Galactic component of the astrophysical neutrino signal and for understanding the propagation of cosmic rays in the interstellar medium. 
}

\titlerunning{Galactic diffuse \gr\ emission with IACTs}
\maketitle

\section{Introduction}

The Milky Way galaxy is the strongest \gr\ source on the sky. Its flux is dominated by the diffuse emission produced by interactions of cosmic ray atomic nuclei and electrons all across the interstellar medium. The reference measurements of the diffuse \gr\ sky are provided by the Fermi Large Area Telescope (LAT; \citet{fermi_diffuse_2012,fermi_template}). More than 10 years of Fermi/LAT exposure have enabled the detection of the diffuse \gr\ flux up to 3 TeV \citep{neronov18,neronov_semikoz19}.  Its spectrum shows a puzzling behaviour extending up to the highest energies as a power law with the slope $dN/dE\propto E^{-\Gamma_\gamma},\ \Gamma_\gamma\simeq 2.4$, which is harder than the slope of the locally measured cosmic ray spectrum ($2.6<\Gamma_{CR}<2.9$) \citep{pdg}, 

This is surprising because the diffuse emission flux in the TeV energy range is expected to be dominated by the pion decay emission from interactions of cosmic ray nuclei. This mechanism results in a \gr\ emission spectrum with a slope close to that of the parent proton and atomic nuclei spectrum \citep{kelner06,kappes}. Therefore, either the average Galactic cosmic ray spectrum is harder than locally measured \citep{malyshev15,yang16}, or the diffuse \gr\ flux has an additional, previously unaccounted for component which provides a sizeable contribution to the overall flux at the highest energies, such as flux from interactions of cosmic rays injected by a nearby source  \citep{Andersen:2017yyg,neronov18,Bouyahiaoui:2018lew,buyo20},
 decays of dark matter particles \citep{Berezinsky:1997hy,Feldstein:2013kka,Esmaili:2013gha,neronov18}, or a large-scale cosmic ray halo around the Galaxy \citep{taylor_aharonian,Blasi:2019obb}.

Pion decay \gr\ emission is always generated together with neutrino emission with approximately equal flux and spectrum \citep{kelner06,kappes}. The Galactic diffuse neutrino emission is not directly detectable by neutrino telescopes  in the GeV-TeV energy range because of the strong background of atmospheric neutrinos. Rather, it only becomes detectable in the energy range above 10 TeV in which IceCube has discovered the astrophysical neutrino signal \citep{icecube_science,icecube_prl}.

The overall flux and spectral characteristics of the IceCube astrophysical neutrino signal \citep{icrc2019} show remarkable consistence with the high-energy extrapolation of the diffuse \gr\ emission spectrum \citep{tchernin14,neronov16,neronov16_1,neronov18}. This might indicate the presence of a sizeable Galactic component of the astrophysical neutrino flux. If the spatial morphology of the Galactic diffuse neutrino and \gr\ emission were well understood and were found to be unchanged with the increase of energy, the search for the Galactic component of the neutrino signal would be possible through fitting of a pre-defined template derived  from the neutrino data \citep{icecube_antares}, a method which currently offers some inconclusive evidence of the signal at a level slightly above $2\sigma$ \citep{icecube19}. Nevertheless, it is likely that the overall pattern of the Galactic diffuse signal changes when the average energy of cosmic rays responsible for the signal production changes from GeV to $>10$~PeV. This is particularly true for the models of  anisotropic diffusion in the  Galactic magnetic field  \citep{Giacinti:2017dgt}. Direct, model-independent identification of the Galactic diffuse \gr$+$neutrino emission signal  would help to clarify the peculiarities of the cosmic ray propagation in the multi-PeV energy range around the knee of the cosmic ray spectrum. 

Such identification is not possible with the IceCube data alone because of the low statistics of the signal: only several tens of neutrinos are detected by IceCube in the energy range above 30~TeV.  One straightforward possibility is to use the \gr\ signal counterpart  to isolate the Galactic neutrino flux component. Indeed, the pion decay \gr s with energies in the 10-100 TeV range could not reach telescopes on the Earth from extragalactic sources because of the effect of absorption on the extragalactic background light \citep{gould,franceschini08}. The multi-TeV diffuse \gr\ flux is therefore coming entirely from the Milky Way.

In what follows we discuss the possibility of detection of the diffuse \gr\ flux in the 10-100 TeV band with the Imaging Atmospheric Cherenkov Telescopes (IACTs) which are conventionally used for statistically robust  observations of TeV \gr\ sources from the ground. We discuss the problem of suppression of the charged cosmic ray background in IACT observations and show that this problem could be overcome so that the diffuse Galactic \gr\ flux is in principle detectable from large portions of the sky with existing  (HESS, MAGIC, VERITAS) and planned (CTA) IACT systems. We also show that the sensitivity of these systems for the diffuse \gr\ flux  is limited by the narrow field of view (FoV). We argue that the IACT technique could be optimised for the measurement of diffuse \gr\ emission and show that a system of small- but wide-FoV IACTs would be able to measure the diffuse \gr\ flux from both low- and high-Galactic-latitude regions on the sky in an energy range overlapping with that of the astrophysical neutrino signal. 

\section{Charged cosmic ray background and its rejection in IACT systems}

Imaging of the Cherenkov light from extensive air showers (EASs)  allows IACT systems to reach large collection areas, $A_{eff}\sim 10^5-10^6$~m$^2$. This is orders of magnitude larger than the effective area of the Fermi/LAT (about $1$~m$^2$) \citep{atwood09} and of the IceCube neutrino telescope ($3-30$~m$^2$ in the 10-100 TeV energy range) \citep{icecube_aeff_cascade,icecube_aeff_muon}. This allows highly statistically significant studies of point or mildly extended \gr\ sources in the TeV range. 

However, the IACT systems are not optimised for measurements of diffuse \gr\ flux. Three obstacles prevent efficient diffuse emission studies. First, the IACTs typically have a narrow          field of view, with a half-opening angle $\Theta_{FoV}$ of  just a few degrees. This provides an angular acceptance of 
\begin{equation}
\Omega\simeq \pi\Theta_{FoV}^2\simeq 2\times 10^{-3}\left[\frac{\Theta_{FoV}}{1.5^\circ}\right]^2
,\end{equation}
 which is orders of magnitude smaller than the FoV of the space-based \gr\ telescope Fermi/LAT $\Omega_{LAT}>2$~sr\footnote{\url{https://fermi.gsfc.nasa.gov/science/instruments/table1-1.html}} or that of the IceCube neutrino telescope, which is $\Omega_{IC}\sim 2\pi$  in the energy range below 1~PeV \citep{icecube_prl,icecube_science,icecube_aeff_cascade}. 
 
 Second, the IACT systems are only able to observe in good weather conditions and moonless nights. This reduces their duty cycle down to approximately $\kappa\sim 10\%$ (compared to the nearly 100\% duty cycle of the space-based \gr\ and ground-based neutrino telescopes). The combination of the first two factors already  significantly reduces the advantage of the large effective collection area, meaning that $\kappa\Omega A_{eff} \sim 10$~m$^2$sr of the IACT systems is comparable to $\kappa\Omega A_{eff}$ of the space-based \gr\ telescope Fermi/LAT and of the neutrino telescope IceCube.

Finally, the space-based \gr\ telescopes and the neutrino telescopes have the capability to efficiently reject charged cosmic-ray-induced background on top of which \gr\ or neutrino signal appears. These telescopes use dedicated systems to veto charged high-energy particles entering the detectors. This is not possible for the IACT systems which use the Earth's atmosphere as a giant high-energy particle calorimeter. The only possibility is to distinguish charged cosmic-ray- and \gr-induced EAS using information on the imaging and timing properties of the EAS signal. Imposing cuts on the imaging and timing characteristics allows  the cosmic ray background to be suppressed by several orders of magnitude. 

\begin{figure}
\includegraphics[width=\linewidth]{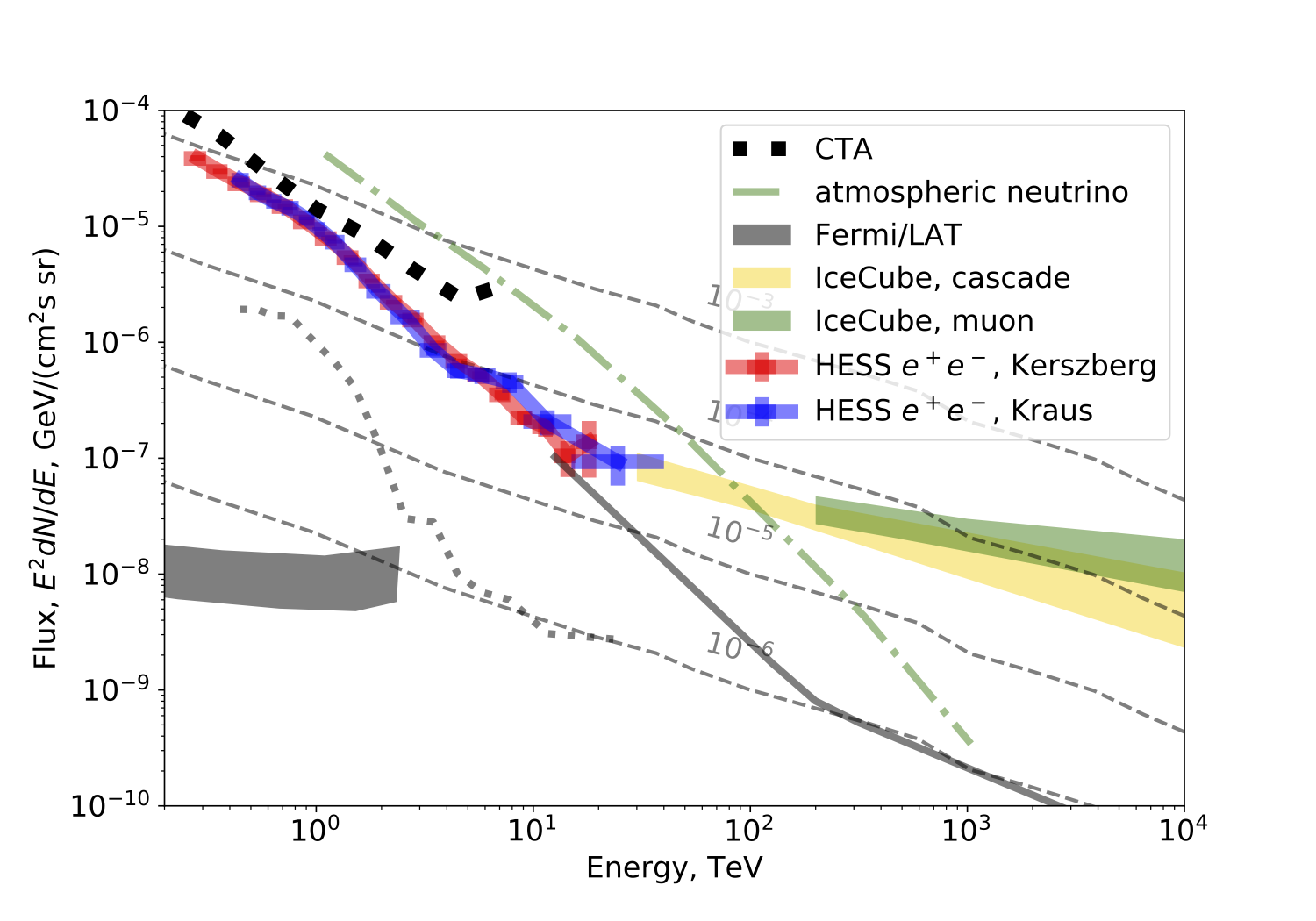}
\caption{Cosmic ray electron spectrum measured by HESS using two different analysis chains reported by \citet{hess_icrc2017,kerszberg,kraus}. Black thick dotted line shows the residual charged cosmic ray background expected in CTA \citep{cta_background}.  Grey shading shows the residual  background level in Fermi/LAT \citep{neronov_semikoz19}.  Dashed grey lines with numerical markers show the flux levels from $10^{-6}$ to $10^{-3}$ of the cosmic-ray all-particle flux. The dotted thin grey curve shows the calculation of the residual cosmic-ray nuclei background  in the HESS electron flux analysis reported by \citet{kraus}. The green dash-dotted line shows the spectrum of the atmospheric neutrino background in neutrino telescopes \citep{icecube_aeff_cascade}. The thick grey  solid line shows the sum of the best-fit model of the cosmic ray electron spectrum and the residual cosmic ray nuclei background suppressed by a factor $10^6$.}
\label{fig:background}
\end{figure}

Figure \ref{fig:background} shows a comparison of the background levels in different types of telescopes. The next-generation IACT array CTA will achieve an efficiency of charged cosmic-ray-background rejection better than $\epsilon_{CR,CTA}\sim 10^{-3}$ \citep{cta_background}. Still, this is much poorer than the $\epsilon_{CR,LAT}\sim 10^{-6}$ efficiency of rejection of the charged cosmic ray background in Fermi/LAT \citep{fermi_igrb,bruel,neronov_semikoz19}. The $10^{-3}$ background suppression factor results in a background flux in the 10-100 TeV energy range which is much higher than the atmospheric neutrino background on top of which the IceCube neutrino telescope detects the astrophysical neutrino signal. 

Better than $10^{-3}$ suppression of the charged cosmic ray background in IACTs could be  achieved by tightening the cuts on the EAS event selection. This is the approach adopted in the analysis aimed at the measurement of cosmic ray electron+positron spectrum with IACTs \citep{hess_electrons_2008,electrons_veritas,electrons_magic,hess_icrc2017,kerszberg,kraus}. Red and blue data points in Fig. \ref{fig:background} show the measurements of the cosmic ray electron spectrum by HESS telescopes \citep{hess_icrc2017,kerszberg,kraus}. These measurements are practically free of the background of cosmic ray atomic nuclei. This efficient rejection of the cosmic ray background is achieved by imposing tight `cuts' on the event selections. Specifically, \citet{hess_icrc2017} adopted the following acceptance criteria \citep{kerszberg}. 
\begin{itemize}
\item events detected by all four HESS telescopes;
\item events with an impact parameter of less than 150~m from the centre of the HESS telescope array;
\item events aligned to within $1.5^\circ$ with the telescope pointing axis;
\item events for which the `mean scaled shower goodness' fit of the shower image with a template of electron or gamma-ray shower image is between $-3$ and $0.6$;
\item events for which the `first interaction point' parameter, corresponding to the distance between the nominal shower-arrival direction and the position of the closest shower image pixel, is between $-1$ and $4$.
\end{itemize}
These cuts are tighter than the `standard' cuts imposed on gamma-ray-like event selections in HESS and other Cherenkov telescope analyses. They reduce the statistical siginificance of the gamma-ray and electron signal while improving the `purity' of the event selection (see \cite{kerszberg} for details).

The upper bound on the residual cosmic ray nuclei background, which was estimated by \citet{kraus} based on Monte-Carlo simulations dedicated to analysis of the electron spectrum, is shown by the dotted curve in Fig. \ref{fig:background}. The statistical significance of EAS events generated by cosmic ray protons and nuclei can be seen to be suppressed by a factor up to $10^{-6}$, i.e. down to a level that is comparable to the residual cosmic ray background in the Fermi/LAT data. 

An essential difference between the background suppression of Fermi/LAT and that of the IACTs is that Fermi/LAT can veto any charged particle background, including cosmic ray nuclei and electrons and positrons, while the best possible background suppression for \gr\ observations with the IACTs can only reject the EASs generated by protons and atomic nuclei. The EASs initiated by primary electrons or positrons are almost indistinguishable from the \gr\ induced showers because the \gr\ initiates EASs via electron--positron pair production, meaning that the electron- and \gr-induced EASs develop in an identical way starting from the second generation of the EAS particles. 

Thus, the minimal possible residual charged cosmic ray background for the diffuse \gr\ flux measurements with IACT is the full cosmic ray electron+positron background shown by the red and blue data points in Fig. \ref{fig:background}. This background has been measured up to an energy of $\simeq 20$~TeV \citep{hess_icrc2017,kerszberg,kraus}. It is accurately modelled by a power law:  $dN/dE\propto E^{-\Gamma_e}$ in the energy range above 2 TeV, with a slope of $\Gamma_e=3.78\pm 0.02(\mbox{stat})_{-0.06}^{+0.17}(\mbox{syst})$ \citep{hess_icrc2017,kerszberg}. High-energy extrapolation of this power law is shown by the grey solid line in Fig. \ref{fig:background}. The steep spectrum of the cosmic  ray electron background leads to a very low level of electron background above 10 TeV energy. The electron flux decreases below $10^{-6}$ of the cosmic ray nuclei flux at an energy of 200~TeV. 

\begin{figure}
\includegraphics[width=\linewidth]{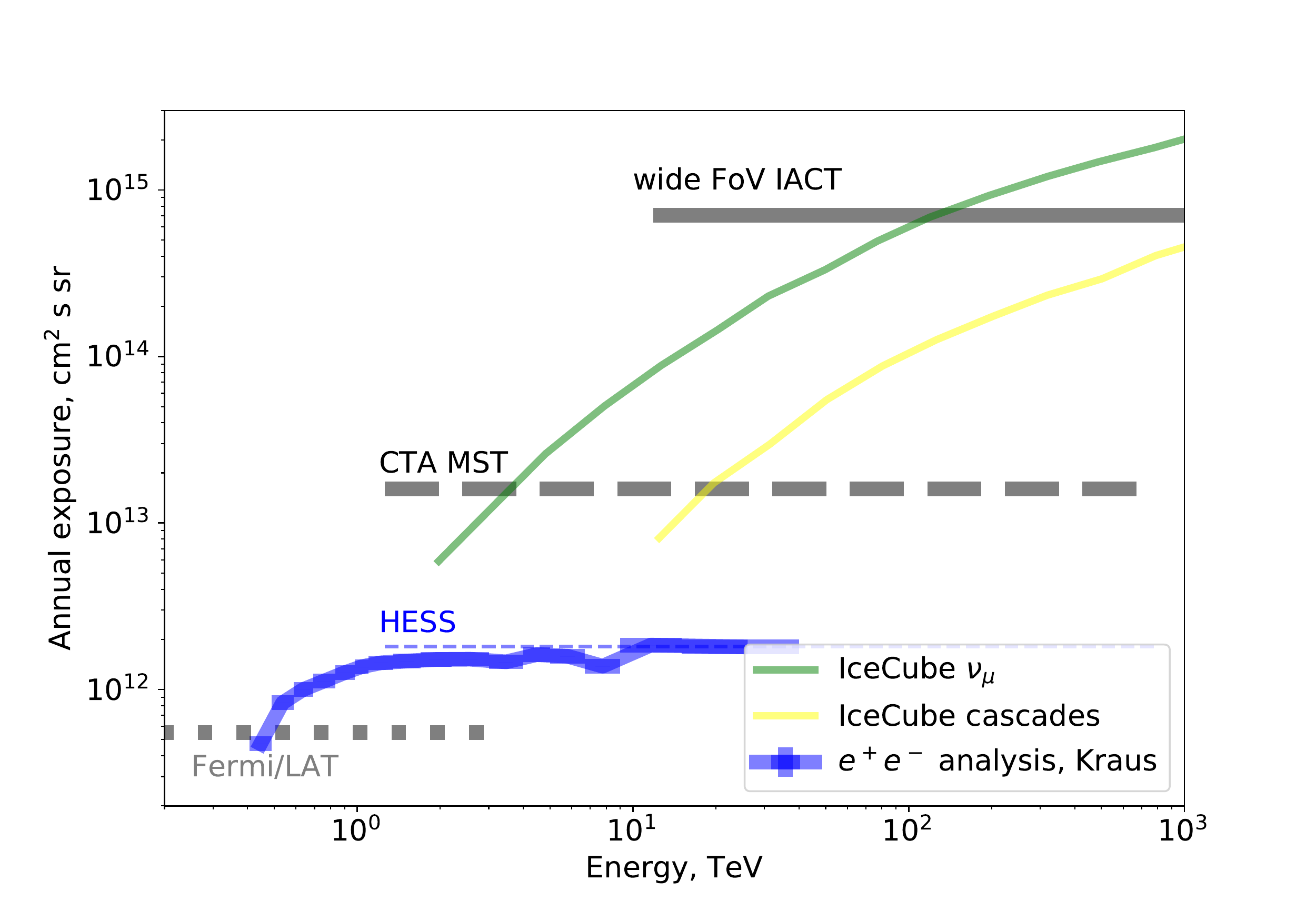}
\caption{Annual exposure of different space- and ground-based  telescopes for diffuse flux measurements. The thick blue histogram shows the exposure of the HESS electron spectrum analysis deduced from the event statistics reported by \citet{kraus}. The thin dashed blue line corresponds to the geometrical exposure limit imposed by the distance and angular cuts adopted in the analysis. The  medium-width yellow and green solid curves show the IceCube exposure in the cascade mode  from \citet{icecube_aeff_cascade} and in the through-going muon neutrino mode  from \citet{icecube_aeff_muon}.  The  thick horizontal short-dashed line shows yearly exposure attainable with Fermi/LAT \cite{atwood09}. The thick horizontal long-dashed and solid lines show the exposure achievable with the MST sub-array of CTA and with a dedicated wide-FoV IACT system for the event selection with the same cuts as in the HESS electron spectrum analysis. }
\label{fig:exposure}
\end{figure}

Better rejection of the charged cosmic ray background in the dedicated electron spectrum analysis  results in a strong reduction of the effective collection area and of the  angular acceptance. Figure \ref{fig:exposure} shows the reduced acceptance $\kappa A_{eff}\Omega$ for the analyses reported by \citet{hess_icrc2017,kraus,kerszberg}. These analyses have reduced the event selection to those EASs hitting the ground within $D=150$~m distance and with the EAS axis misaligned with the telescope pointing direction by at most $1.5^\circ$. This results in a geometrical acceptance corresponding to the thin dashed horizontal line in Fig. \ref{fig:exposure} (under assumption of the duty cycle $\kappa=0.1$). Blue data points in Fig. \ref{fig:exposure} show the acceptance derived from the cosmic ray electron event statistics reported by \citet{kraus}. The acceptance of the electron spectrum analysis can be seen to correspond to the geometrical acceptance above 1 TeV energy.

\section{Sensitivity of existing and planned IACT systems for the Galactic diffuse \gr\ flux}

The low level of residual charged cosmic ray background achievable with the IACT technique opens the possibility of background-free measurements of diffuse Galactic \gr\ emission in the energy band above that attainable for Fermi/LAT and possibly overlapping with that of the astrophysical neutrino signal. However,  the small geometrical acceptance of the `minimal cosmic ray background' event sample leads to a low statistical significance for the signal even for the  electron background (and, respectively, of the diffuse \gr\ signal on top of this background) in the energy range above 10 TeV. Less than ten events are reported within a 1000 hr exposure which corresponds to approximately one year of observations with the duty cycle $\kappa\sim 10\%$  \cite{hess_icrc2017,kerszberg,kraus}. This limits the possibility of measuring the diffuse \gr\ flux from the Galaxy with existing IACT systems: HESS, MAGIC, and VERITAS. 

\begin{figure}
\includegraphics[width=\linewidth]{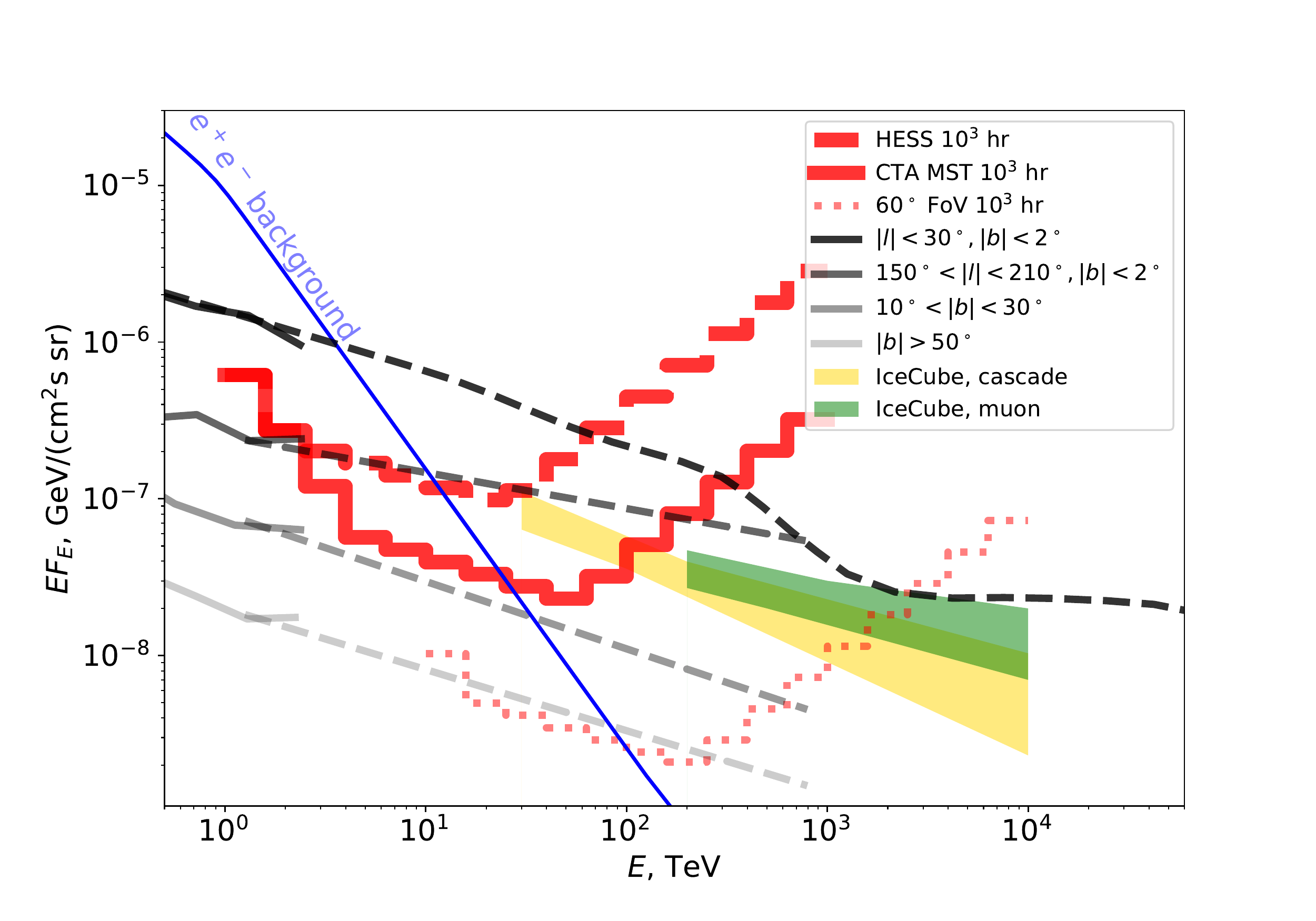}
\caption{Sensitivity of CTA  (red thick solid line) and HESS (red thick dashed line) telescopes for detection of the Galactic diffuse \gr\  emission. Green and yellow butterflies show the spectrum of the astrophysical neutrino signal detected in the  cascade and through-going muon detection modes  \citep{icrc2019}. Light-to-dark grey solid lines show the flux levels of diffuse \gr\ emission from different parts of the sky measured by Fermi/LAT, as reported by \citet{neronov_semikoz19}. Dashed lines of the same colour show high-energy power law extrapolation of the Fermi/LAT measurements.  The Galactic Ridge signal (the darkest grey line) power law is modified by the effect of pair production on cosmic microwave background assuming a distance of 8 kpc.  The dotted light red curve shows the sensitivity which could be achieved with a dedicated array of telescopes with a small and wide FoV. }
\label{fig:cta}
\end{figure}

The red dotted line in Fig. \ref{fig:cta} shows the calculated differential sensitivity of HESS for the diffuse \gr\ flux.  This calculation adopts the standard conventions used for differential sensitivity calculations. The minimal detectable flux is calculated per energy bin, assuming logarithmic energy binning with five bins per decade. In each energy bin, the minimal detectable signal $S$ is determined by the statistics of the background events $B$:
\begin{equation}
S_{min}=\mbox{max}\left\{
\begin{array}{l}
0.1B\\
3\sqrt{B}\\
3
\end{array}
\right\}
.\end{equation}
The minimal signal is not less than $10\%$ of the background level, or exceeds the background fluctuations by at least $3\sigma$, but in any case the signal is required to be not less than three counts even if the background is negligibly small. 

The sensitivity of potential  HESS measurements of the diffuse \gr\ flux is limited by the statistical  fluctuations of the background level starting from the energy $\simeq 2$ TeV for a 1000 hr exposure (comparable to that of the HESS electron spectrum analysis \citep{hess_icrc2017,kerszberg,kraus}. The sensitivity is limited by the signal statistics above 30 TeV. 

Figure \ref{fig:cta} shows the levels of the Galactic diffuse \gr\ flux from different parts of the sky, which could be inferred from high-energy power law extrapolation of the Fermi/LAT measurements up to 3 TeV, reported by \citet{neronov_semikoz19}. Extrapolation of the flux from the Galactic Ridge up to PeV energies  includes the effect of attenuation of the \gr\ flux by pair production on low-energy photon backgrounds, mostly on cosmic microwave background. Attenuation of the infrared interstellar radiation field background is less important \citep{isrf}, but this latter is still visible as a minor feature superimposed on the power law spectrum somewhat below 100 TeV.  Comparison of the flux levels with the HESS  1000 hr exposure sensitivity shows that the regions of diffuse emission along the Galactic Plane, from the Galactic Ridge, $-30^\circ<l<30^\circ$, up to the outer Galaxy, $150^\circ<l<210^\circ$, are detectable. On the contrary, the diffuse flux from the mid-latitude region, $10^\circ< |b|<30^\circ$, and from the high Galactic latitude regions, $|b|>50^\circ$, are not accessible to the long HESS exposure.  The diffuse emission flux estimate is made taking into account the fact that the total flux is composed of the diffuse and resolved source contributions. The resolved source contribution is estimated from the count statistics within $0.5^\circ$ circles around sources from the Fermi catalogue (see \citet{neronov_semikoz19} for details).

\begin{figure}
\includegraphics[width=\linewidth]{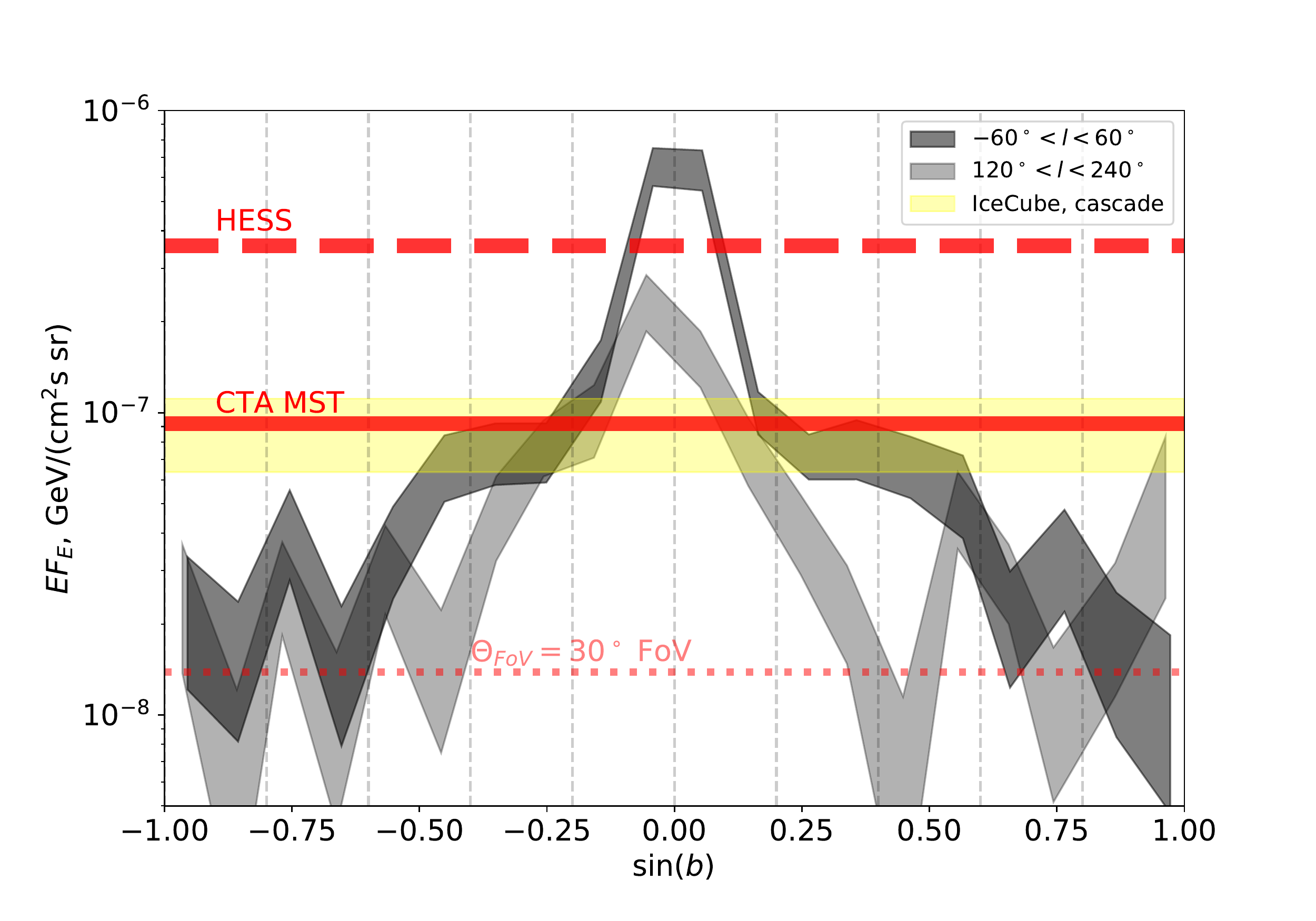}
\caption{Sensitivities of HESS, CTA/MST and a hypothetical $30^\circ$ FoV IACT system for measurements of the Galactic latitude profiles of diffuse emission (100 hr exposures per bin, each of $\Delta \sin(b)=0.1$   in width, 1000 hr total exposure) at 30~TeV reference energy. Yellow horizontal shading shows the level of the astrophysical neutrino flux measured by IceCube in cascade mode \citep{icrc2019}. Light and dark grey shaded bands show an extrapolation of the Fermi/LAT measurements at 1 TeV, assuming the spectral slope $\Gamma_\gamma=2.4$ \citep{neronov_semikoz19}. }
\label{fig:profiles}
\end{figure}

This is illustrated in a different way in Fig. \ref{fig:profiles} which shows a comparison of the Galactic latitude profiles of the diffuse \gr\ flux extrapolated from 1 to $30$~TeV energies assuming a power-law spectrum with a slope of $\Gamma_\gamma=2.4$ with the sensitivity of HESS at this energy. To measure the Galactic latitude profile of the diffuse flux in the inner or outer Galaxy parts of the sky, an IACT would need to `scan' different Galactic latitude regions with its narrow FoV. Assuming that the full year-long (1000 hr) exposure is divided into ten 100~hr intervals  for each of the Galactic latitude bins of the width $\Delta\sin(b)=0.1$, only the inner Galaxy exposure toward the Galactic Plane would give a detection with HESS. 

Such a detection has already been reported by \citet{hess_diffuse} as a byproduct of the inner Galactic Plane survey performed by HESS. The analysis of \citet{hess_diffuse} did not apply tighter cuts on event selection to reduce the residual cosmic ray background level. Instead, it adopted the assumption that certain regions of the sky (above Galactic latitude $|b|=1^\circ$) are free from diffuse \gr\ flux and contain only the residual cosmic ray background flux. These regions were used to estimate the residual cosmic ray background level. This generically results in over-subtraction of the background, because the regions used for the background estimate do contain weaker diffuse \gr\ flux (see \citet{neronov_semikoz19} for details).  This results in a  mismatch between the background-free measurements of the diffuse flux by Fermi/LAT and the lower bounds stemming from HESS analysis in some sky regions by up to a factor of two. An improvement of the \citet{hess_diffuse} could be achieved using the method proposed in this paper: tightening of the cuts on event selection to achieve background-free detection of the diffuse emission.

This example shows the generic potential of the diffuse \gr\ flux measurements by IACT systems: they do not require dedicated exposures, but rather could be obtained as a byproduct of regular observational campaigns which nevertheless cover different Galactic longitude and latitude regions. Thus, the use of an approximately 1000 hr (one-year long span of observations) exposure adopted in our sensitivity calculations appears reasonable in this context.

Limitations of the HESS `minimal cosmic ray background' mode sensitivity point to possible ways of improvement of IACT configurations for the diffuse flux measurements:
\begin{itemize}
\item deploying a larger number of telescopes in an IACT array to increase the collection area $A_{eff}$ and/or
\item implementing  telescopes in the IACT array with a wider FoV  to extend the angular acceptance $\Omega$.
\end{itemize}
Both improvements will be implemented in the CTA observatory. The Medium Size Telescopes (MST) of CTA will have dish diameters comparable to those of HESS telescopes, but will have a wider FoV of  $3.75^\circ$ in radius (the telescopes of HESS have a FoV of $2.5^\circ$). The electron analysis of HESS was limited to the effective FoV $\Theta_{FoV,HESS,e}=1.5^\circ$ \citep{hess_icrc2017,kerszberg,kraus} to achieve the highest rejection of nuclear cosmic rays. Assuming that the same $1^\circ$ margin could be used for the MST `electron' analysis, the CTA MST telescopes will have an effective FoV of $\Theta_{FoV,MST,e}=2.75^\circ$. The dense geometrical arrangement of the MST telescopes in the CTA North sub-array  will allow us to achieve a collection area that is a factor of $2.6$ larger than that of the HESS array in the electron EAS detection mode (event selection in this mode has imposed an EAS impact distance cut of $D<150$~m from the centre of the HESS array where telescopes are arranged in a grid of squares, each square being 120m in length and width). The 15 MST telescopes of CTA North will be arranged in a grid with a similar telescope spacing\footnote{https://www.cta-observatory.org/about/array-locations/la-palma/}. The combination of the increased collection area and extended FoV will provide an order-of-magnitude gain in geometrical acceptance of the CTA North MST sub-array compared to HESS, as shown in Fig. \ref{fig:exposure}. 

The MST sub-array will provide better acceptance than the Large Size Telescope (LST) sub-array, which will have smaller FoV telescope units and a smaller number of telescopes. The Small Size Telescope (SST) sub-array foreseen for the Southern CTA site will possibly not provide the `minimal charged cosmic ray background' configuration because of the large spacing between the telescopes, much larger than the distance cut of $D<150$~m  imposed in the HESS electron spectrum analysis (which guarantees a maximum distance of 60 m  from the shower impact point to the nearest telescope and assures that  and several telescopes are providing the images from within the Cherenkov light cone footprint). 

The order-of-magnitude improvement in the acceptance of the CTA MST sub-array will provide a significant improvement of sensitivity for the detection of the diffuse \gr\ flux, as can be seen from Fig. \ref{fig:cta}.  A 1000 hr exposure (corresponding to one year of data) will be sufficient for detection of the diffuse emission at the level of the astrophysical neutrino flux  in the energy range overlapping with that of IceCube measurements. 

Figure \ref{fig:profiles} shows that  if the 1000 hr year-long exposure of CTA MST is distributed over different Galactic latitudes, with $\sim 100$~hr per $\Delta\sin(b)=0.1$ Galactic latitude bin, CTA will still not be able to fully map the diffuse flux at the level of the IceCube neutrino flux over the full sky; it will only be able to firmly detect the Galactic emission up to the latitudes $|b|<5^\circ$  ($|\sin(b)|<0.1$). Several years of exposure would be needed to achieve detections up to $ 
|b|\sim 30^\circ $.

\section{Discussion: Possible optimisation of the  IACT technique for the diffuse flux search}

The results of the previous sections show that the IACT systems are suitable for detection of the diffuse Galactic \gr\ emission in the energy range above the current limit at 3~TeV. These measurements could be derived as a byproduct of regular observation campaigns, by choosing the `minimal charged cosmic ray background' event selections similar to those produced for the analysis of the cosmic ray electron spectrum. 

\begin{figure}
\includegraphics[width=\linewidth]{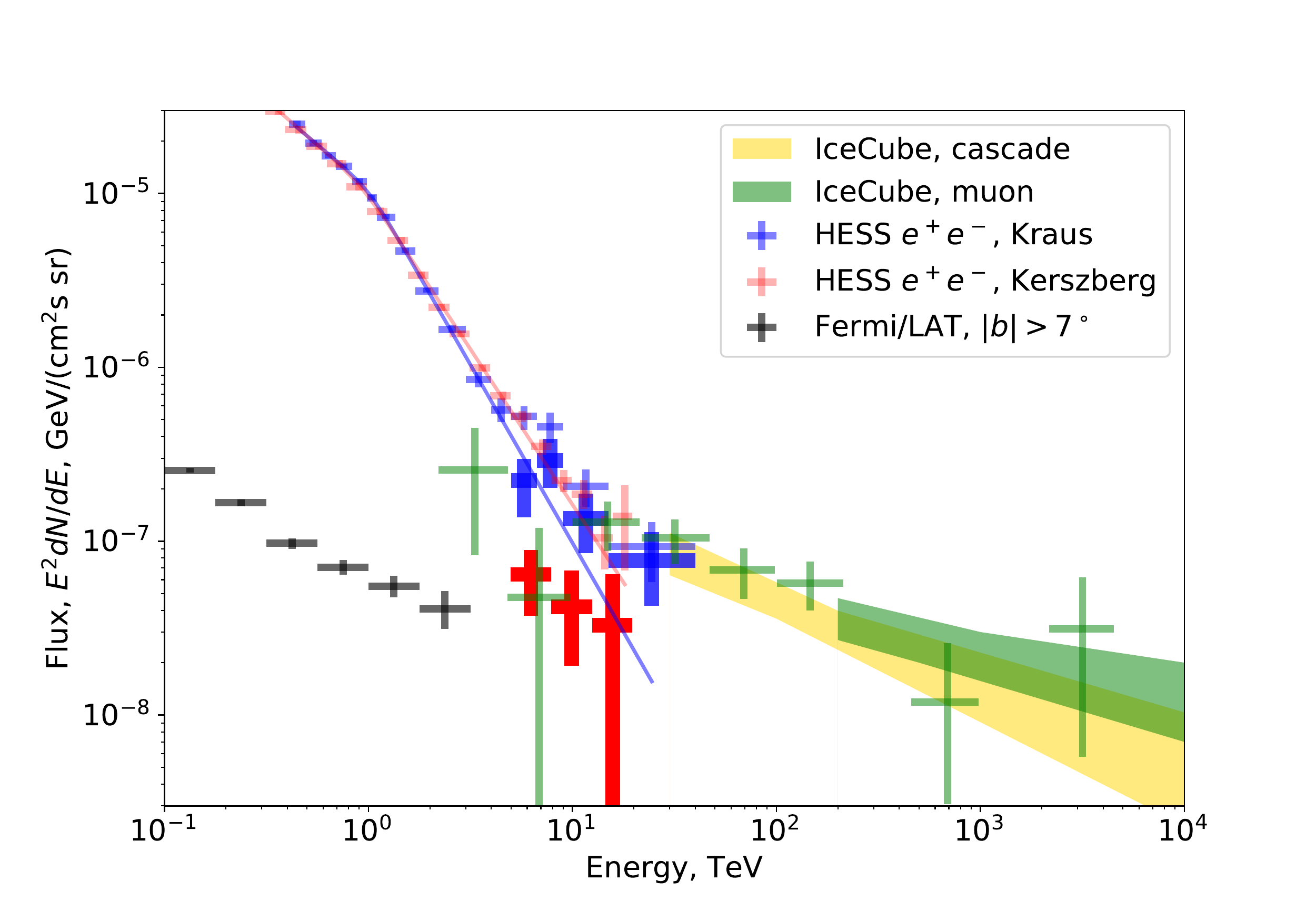}
\caption{ High-energy excesses (blue and red data points) over power-law models (light blue and red lines) of the HESS electron spectrum measurements by \cite{kraus} and \citet{hess_icrc2017,kerszberg} (shown by thin light blue and red data points). The thin dark grey data points below 3 TeV show Fermi/LAT measurements of the Galactic diffuse flux from the Galactic latitude region $|b|>7^\circ$, extracted in the same way as in the analysis of \cite{neronov_semikoz19}. Yellow and green butterflies show IceCube astrophysical neutrino flux spectra derived from the analysis of cascade and through-going muon neutrino \citep{icrc2019} event selections. Green data points show the neutrino spectrum reported by \citet{icecube_icrc2017}.  }
\label{fig:hess_fermi}
\end{figure}

Among the existing IACT systems, HESS has the largest FoV and hence provides the highest sensitivity for the diffuse \gr\ flux. Its electron spectrum analysis technique could be directly used to obtain a measurement of the diffuse Galactic \gr\ flux above energies of several TeV  in the Galactic Ridge ($|l|<30^\circ$, $|b|<2^\circ$) region; see Figs. \ref{fig:cta} and \ref{fig:profiles}. A multi-year exposure of HESS could be sufficient for detection of the diffuse emission even from  regions of higher Galactic latitude. This is illustrated in Fig. \ref{fig:hess_fermi}, where thick blue and red  data points show mild high-energy excesses of the electron spectra derived by \citet{kraus,hess_icrc2017,kerszberg} over broken power-law models derived from the fits to lower energy data. Comparing these excesses with the level of the IceCube astrophysical neutrino flux and with the  Fermi/LAT diffuse sky flux from the region $|b|>7^\circ$ (corresponding to the data selection criterium of HESS analysis \citep{hess_icrc2017,kerszberg}) we find that the overall excess flux levels are comparable to expected diffuse \gr\ flux from the sky region covered by the HESS analysis (the quoted systematic error on the electron flux is $\Delta\log(EF_E)\simeq  0.4$). The overall excesses within $805$ and $1186$~hr of HESS exposures \citep{kraus,kerszberg}  are at the levels of $>4\sigma$ for the analysis of \citet{kraus} and  $1.7\sigma$ for the analysis of \citet{kerszberg}. A factor-of-ten longer exposure (which is potentially already available with HESS) could reveal a higher significance excess at the level of up to $5\sigma$. Such an excess is predicted in a range of theoretical models including  interactions of cosmic rays injected by a nearby source  \citep{Andersen:2017yyg,neronov18,Bouyahiaoui:2018lew}
 or decays of dark matter particles \citep{Berezinsky:1997hy,Feldstein:2013kka,Esmaili:2013gha,neronov18} or a large-scale cosmic ray halo around the Galaxy \citep{taylor_aharonian,Blasi:2019obb}.

The geometric aperture $\kappa A_{eff}\Omega$ of the existing IACT arrays is still too small for high-quality mapping of the diffuse emission from higher Galactic latitudes and from the outer Galactic disc.  A crucial improvement of the aperture will be provided by the CTA and in particular by its Medium Size Telescope (MST) sub-array. The CTA measurements using the `minimal charged cosmic ray background' event selection (with the background dominated by the cosmic ray electrons) will allow, for the first time, direct probing of the diffuse \gr\ flux level comparable to the level of the IceCube astrophysical neutrino signal in the energy range overlapping that of the IceCube measurements. This will be an important milestone for direct identification of the Galactic component of the astrophysical neutrino flux. 

However, the sensitivity level achievable with the CTA within one year of operation will still be only marginally sufficient for the mapping of the signal with the flux at the level of the IceCube neutrino flux in different sky regions. This is clear from Fig. \ref{fig:profiles} where a 1000 hr CTA exposure is supposed to be split into ten 100 hr exposures at different Galactic latitudes. Approximately $ 10$~yr of operation time (a  10 000 hr exposure split onto ten 1000 hr exposures) will rather be required to test a model in which a new Galactic flux component distributed over a large range of Galactic latitudes (such as a local source or a dark matter decay signal) is responsible for the astrophysical neutrino signal. Nevertheless, this multi-year exposure, which does not require a dedicated observation time and will be a byproduct of the regular CTA observations,  has the potential to  provide direct proof of the existence of such a new Galactic flux component. 

The IACT observation technique could in principle be explicitly optimised for the increase of the acceptance $\kappa A_{eff}\Omega$  in the energy range of the IceCube astrophysical neutrino signal, $E>30$~TeV. The $E>30$~TeV EASs hitting the ground within $D<150$~m distance (comparable to the radius of the Cherenkov light pool footprint) are detectable with telescopes of apertures much smaller than that of HESS. Figure \ref{fig:exposure} shows that the acceptance of the HESS telescopes in the electron spectrum analysis configuration is equal to the geometrical acceptance already starting from $E\simeq 1$~TeV energy. A $E\sim 30$~TeV EAS produces a signal comparable to that of a 1~TeV shower in HESS (telescope dish size $D_{tel}\simeq 12$~m) already in a telescope with $D_{tel}\simeq 2.5$~m. Therefore, an IACT system composed of small 2.5~m diameter telescopes arranged similarly to  the HESS telescopes could already achieve the geometrical aperture equivalent to HESS at 30~TeV  in the `minimal charged cosmic ray background' mode. 

An obvious advantage of such a   system of smaller telescopes is that construction and operation costs are significantly reduced compared to those for HESS or the CTA MST sub-array. This opens the possibility for extension of the angular acceptance $\Omega$, for example by using a wide FoV optical system. Examples of $2.5$~m class, wide-FoV telescopes are provided by EUSO space-based fluorescence telescope\footnote{http://jem-euso.roma2.infn.it} for which Schmidt telescope optics or a refractor telescope equipped with Fresnel lenses are considered \citep{euso,euso1,euso2}.  All EUSO telescope configurations implement an optical system which achieves the aperture $D_{tel}\simeq 2.5$~m and provides a point spread function of  $\theta_{psf}\simeq 0.1^\circ$ across a very wide FoV: $\Theta_{FoV}\simeq 30^\circ$ . The Schmidt optics, consisting of a spherical mirror and a Fresnel corrector plate at the telescope entrance, were also considered as a possible option for the wide FoV IACT by \cite{mirzoyan}. A very wide FoV could also be achieved via coverage of a wide solid angle with the overlapping FoVs of IACTs sub-arrays. A total $\Theta_{FoV}\simeq 30^\circ$ is achievable with one of the approaches outlined above, providing the geometrical acceptance shown by the grey thick solid horizontal line in Fig. \ref{fig:exposure}. This is possible already with the relatively small effective area with the cut on the maximal EAS impact distance $D=150$~m. This geometrical acceptance is almost three orders of magnitude larger than that of the HESS telescopes in the electron spectrum analysis mode.

Such an increase of acceptance could lead to a crucial improvement of sensitivity for the measurement of the diffuse \gr\ flux. Re-calculation of the differential flux sensitivity using the same approach as described in the previous section leads to the result shown by the light-red dotted line in Fig. \ref{fig:cta}. This figure shows that a year-long operation ($1000$~hr exposure) will provide sufficient sensitivity  for detection of diffuse \gr\ emission even from regions of high Galactic latitude:  $|b|>50^\circ$. This is also shown in Fig. \ref{fig:profiles} where the 1000 hr exposure is divided into ten 100 hr exposures in different Galactic latitude bins. Both the signals with nearly isotropic sky flux patterns (the yellow band of the astrophysical neutrino signal) and with strong anisotropy toward the Galactic Plane (grey bands of the extrapolated Fermi/LAT measurements) could be explored with the wide-FoV small telescope system.  

\section{Conclusions}

Here, we show that the IACT technique could be used measure the diffuse Galactic \gr\ flux in the energy range 10-100 TeV overlapping with the range of the IceCube measurements of the astrophysical neutrino flux. Such measurements are possible using the event selections designed for measurement of the cosmic ray electron spectrum. These event selections are free of the background of proton and heavier nuclei cosmic rays. 

We show that the existing decade-long exposures by the current generation of \gr\ telescopes could be used to look for evidence of the existence of the \gr\ counterpart of the IceCube astrophysical neutrino flux. Definitive identification of the Galactic component of the neutrino flux could be achieved by the MST sub-array of the CTA. 

We also propose that the IACT technique could be optimised specifically for the study of the diffuse Galactic \gr\ flux in the IceCube energy range. Such optimisation could be achieved with  an IACT system of telescopes with relatively small (2.5 m size) but wide FoVs ($\Theta_{FoV}\simeq 30^\circ$) using Schmidt or refractor telescope optics. This optimised system would enable highly statistically significant mapping of the 10-100 TeV diffuse Galactic \gr\ flux across the entire sky. 

\bibliographystyle{aa}
\bibliography{Diffuse_gamma_HESS}
\end{document}